\documentclass[prl,floats,twocolumn,aps,superscriptaddress,showpacs]{revtex4}
\usepackage{amsmath}
\usepackage{graphicx}

\newcommand{\beq}{\begin{equation}}
\newcommand{\eeq}{\end{equation}}
\newcommand{\be}{\begin{equation}}
\newcommand{\ee}{\end{equation}}
\newcommand{\bea}{\begin{eqnarray}}
\newcommand{\eea}{\end{eqnarray}}
\newcommand{\barr}{\begin{array}}
\newcommand{\earr}{\end{array}}

\begin{document}
\title{ {\bf Noise activated granular dynamics}}

\author{Fabio Cecconi}
\author{Andrea Puglisi}
\affiliation{Dipartimento di Fisica, 
Universit\`a ``La Sapienza'', P.le A. Moro 2,~ 
I-00185 Rome, Italy and\\
INFM Center for Statistical Mechanics and Complexity, Italy}         
\author{Umberto Marini Bettolo Marconi}
\affiliation{Dipartimento di Matematica e Fisica, Universit\`a di Camerino,
Via Madonna delle Carceri, I-62032 , Camerino, Italy and
INFM, Unit\`a di Camerino}
\author{Angelo Vulpiani}
\affiliation{Dipartimento di Fisica, 
Universit\`a ``La Sapienza'', P.le A. Moro 2,~ 
I-00185 Rome, Italy and\\
INFM Center for Statistical Mechanics and Complexity, Italy}         

\begin{abstract}

We study the behavior of two particles moving 
in a bistable potential, colliding inelastically with each other
and driven by a stochastic heat bath. The system has the tendency to
clusterize, 
placing the particles in the same well at low drivings, and to fill 
all of the available space at high temperatures. We show that the hopping
over the potential barrier occurs following the Arrhenius rate, 
where the heat bath temperature is replaced by the granular temperature.
Moreover,
within the clusterized ``phase'' one encounters two different scenarios:
for moderate inelasticity, the jumps from one well to the other involve one 
particle at a time, whereas for strong inelasticity the two particles 
hop simultaneously.

\end{abstract}
\maketitle


Granular gases~\cite{gases}, i.e. assemblies of inelastic particles
losing a little kinetic energy at each collision, exhibit a variety of
complex behaviors, such as clustering~\cite{goldhirsch}, spontaneous
formation of vortices~\cite{bal}, lack of energy 
equipartition~\cite{Ioepug}, 
non-Maxwellian velocity distributions~\cite{Nogauss}, and so on, which 
provide new challenges to statistical mechanics. 
Analogies between standard
condensed matter and granular matter remain a valuable 
route to improve our knowledge
of the latter~\cite{jaeger}. As an instance, we mention 
the analogy between the granular temperature, $T_G$,
defined as the average kinetic energy per grain, 
and the temperature of molecular gases.  
It has been observed that the equality of the granular temperatures
of two granular gases is not the condition for thermal equilibrium
between them \cite{Ioepug}. 
This is not too surprising since it reflects the
non-equilibrium nature of granular systems. Nevertheless, 
one can ask whether $T_G$ possesses other useful properties
of the thermodynamic temperature, such as that of controlling
the rates of activated processes and the direction of energy
fluxes. The answer is relevant in the construction of
the hydrodynamics of granular systems.

The present work is inspired by the experiment probing the behavior of
vibrated sand in a vertical box made up of two identical compartments
communicating through a small orifice located at a certain height. One
observes that for vigorous shaking the two halves are equally
populated, whereas below a critical driving intensity the symmetry is
broken~\cite{experiment}. The existing theoretical explanations are
based on hydrodynamic descriptions. According to
Refs.~\cite{Eggers,Brey,Lohse}, the behavior can be captured by a
phenomenological mesoscopic flux model, depending on a parameter which
is a function of the inelasticity, the driving intensity and the number
of particles. One assumes the existence of a stationary state and,
focusing on slow variables, neglects the role of temporal and spatial
fluctuations.
  
Our treatment, instead, represents a shift from the hydrodynamic 
to the statistical mechanical level, in which 
quantities such as temperature and density fluctuations are obtained in 
terms of the microscopic coordinates of the particles and 
their interactions. 
In a simplified model, we relate the crossover from the equally 
populated phase to a broken symmetry phase to the existence of two different 
kinds of thermally activated processes.
This is a strong indication that the kinetic temperature
is a parameter characterizing, not only velocity fluctuations, but also
the activation dynamics across energy barriers. 

Futhermore, the simplicity of the model allows us to make explicit the
dependence of the kinetic temperature on the relevant parameters 
of the model.

Our model is closely connected to that, proposed over sixty years
ago by H.A. Kramers~\cite{Kramers,Hangi}, describing a reaction occuring via
a thermally activated barrier crossing. We consider two inelastic
hard rods (the simplest granular gas) 
bound to move on a line in the presence of a
bistable external potential $U(x)=-ax^2/2+b x^4/4$, 
(mimicking the compartimentalized
box). The particles are coupled to a bath which exerts
upon them a velocity dependent friction and a random force. 
In the absence of collisions the particles evolve according to: 

\begin{equation}
M \frac{d^2 x_i}{dt^2}=-M \gamma\frac{d x_i}{dt} - U'(x_i)+\xi_i(t)
\label{kramers}
\end{equation}
where, prime indicates the spatial derivative, $x_i$ ($i=1,2)$ represents
the position of particles and $\gamma$ is a friction coefficient.
The Gaussian noise $\xi_i(t)$, with $<\xi_i(t)>=0$ and $<\xi_i(t)\xi_j(s)>=2 M K_b \gamma
T_b\delta_{ij} \delta(t-s)$, describes the exchange of energy with the
surroundings \cite{tradition} schematized by a heat-bath of temperature
$T_b$.
In the following we set $M=1$, $K_B=1$, $a=1.5$, $b=0.05$
and $\gamma=0.1$. The relevant parameters are: the positions of minima
$x_{min}=\pm\sqrt{a/b}=L/2$, the location of the maximum,
$x_{max}=0$, and the energy barrier,
$\Delta U=a^2/4b=11.25$, together with the curvatures
$\omega_{min}^2=U''(x_{min})=2a$ and $\omega_{max}^2=-U''(x_{max})=a$.

Upon taking into account collisions, the velocities of
particles change instantaneously when their separation $x_2-x_1$
equals the hard core diameter, $d=1$, according to the rule:
\begin{equation}
 v_i'  =  v_i-
\frac{ 1+r}{2}(v_i-v_j) 
\label{collision}
\end{equation}
where $0 \leq r \leq 1$ is a restitution coefficient and $i\neq
j=1,2$.  Let us recall that in the non-interacting case (no
collisions), each particle on the average sojourns a time given by
\begin{equation}
\tau=A \exp(\Delta U/T_b), 
\label{kram}
\end{equation} 
where the prefactor $A$ has been calculated by
many authors since the work of Kramers \cite{Hangi}.

\begin{figure}[htb]
\begin{center}
\includegraphics[clip=true,width=\columnwidth, keepaspectratio]
{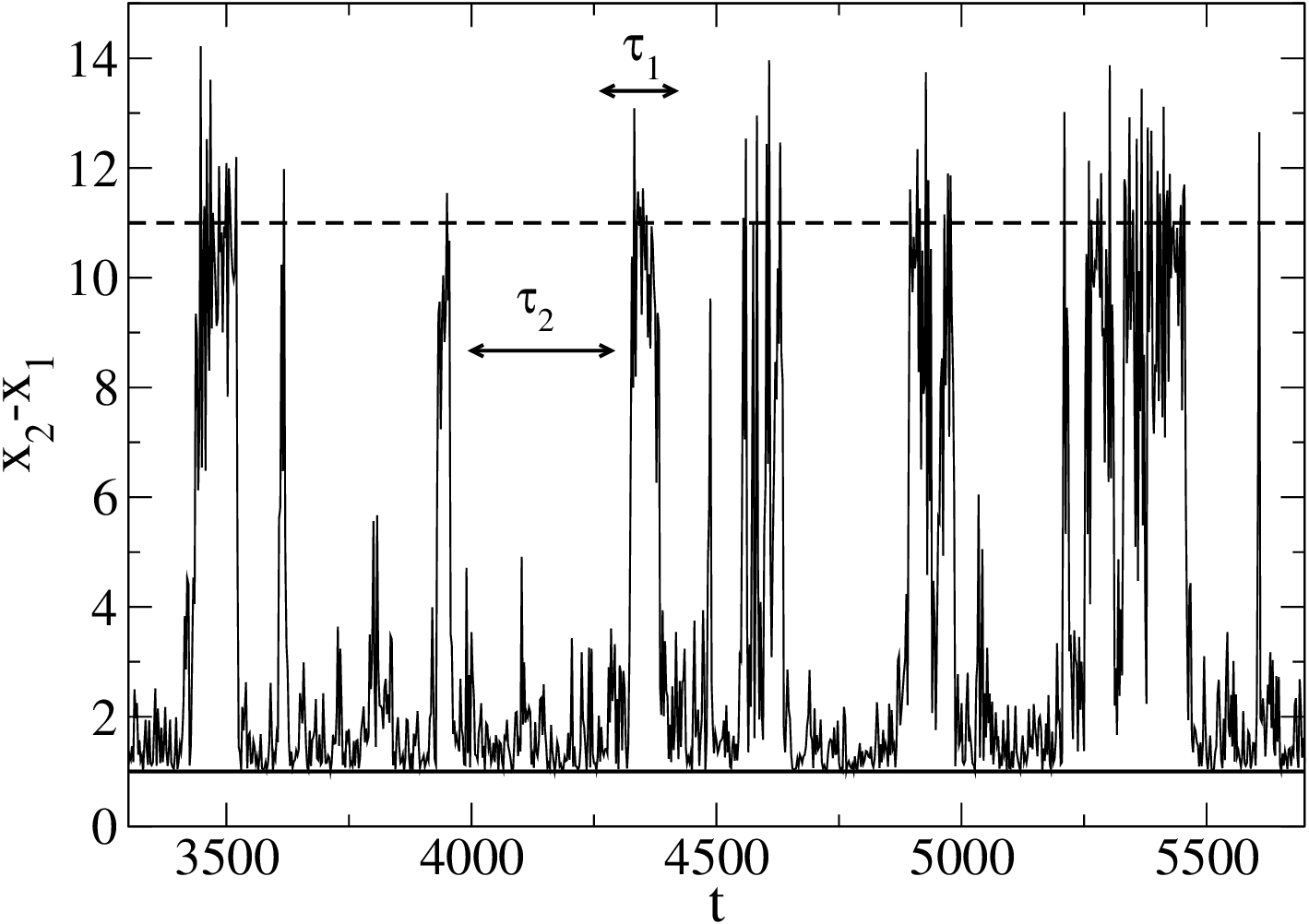}
\end{center}
\caption
{Relative distance $x_2-x_1$ as a function of time for a system with $r=0.9$ 
and $T_b=4.0$.  
The solid line indicates the diameter of the rods, 
the dashed marks the well 
separation $L\simeq 10.95$.}
\label{fig_xrel}
\end{figure}

The basic phenomenology of the model is illustrated in Fig.~\ref{fig_xrel}.
The relative distance, $y=x_2-x_1$, between particles fluctuates in time
showing time intervals of average lifetime
$\tau_2$, when they are confined to the same well ($y\sim d$)
alternated with intervals, of average
lifetime $\tau_1$, when they sojourn in separate wells ($y \sim L$).
Thus, the system behaviour shows the existence of 
two different time scales $\tau_1$ and $\tau_2$ 
employed to characterize two regimes.

We shall see that, as the driving intensity, $T_b$, or  
inelasticity, $r$, are varied, the system undergoes a crossover from the 
regime in which particles are far apart most of the time 
(i.e. $\tau_1 > \tau_2$), to a clusterized regime characterized by 
$\tau_1<\tau_2$. 
Moreover, we shall show that the
dependence of $\tau_2$ and $\tau_1$ on the model parameters can be captured
by a simple extension of formula~(\ref{kram}), replacing $T_b$ by the 
two different kinetic temperatures 
\begin{eqnarray}
T_2 &=& \lim_{t\to\infty} 
\frac{1}{2t_2} \int_{0}^t ds\bigg(v^2_1(s) + v^2_2(s)\bigg) 
\Theta[x_1(s) x_2(s)]  \\
T_1 &=& \lim_{t\to\infty} 
\frac{1}{2t_1} \int_{0}^t ds\bigg(v^2_1(s) + v^2_2(s)\bigg) 
\Theta[-x_1(s) x_2(s)] 
\end{eqnarray}
where $t_1$ and $t_2$ are the times the particles spend, during $[0,t]$, 
in different wells or in the same one respectively. 
The temperature $T_2$ represents
the velocity variance conditioned to the fact that the two particles
belong to the same well, whereas $T_1$ is the same quantity when these
move in different wells. 

\begin{figure}[htb]
\begin{center}
\includegraphics[clip=true,width=\columnwidth, keepaspectratio]
{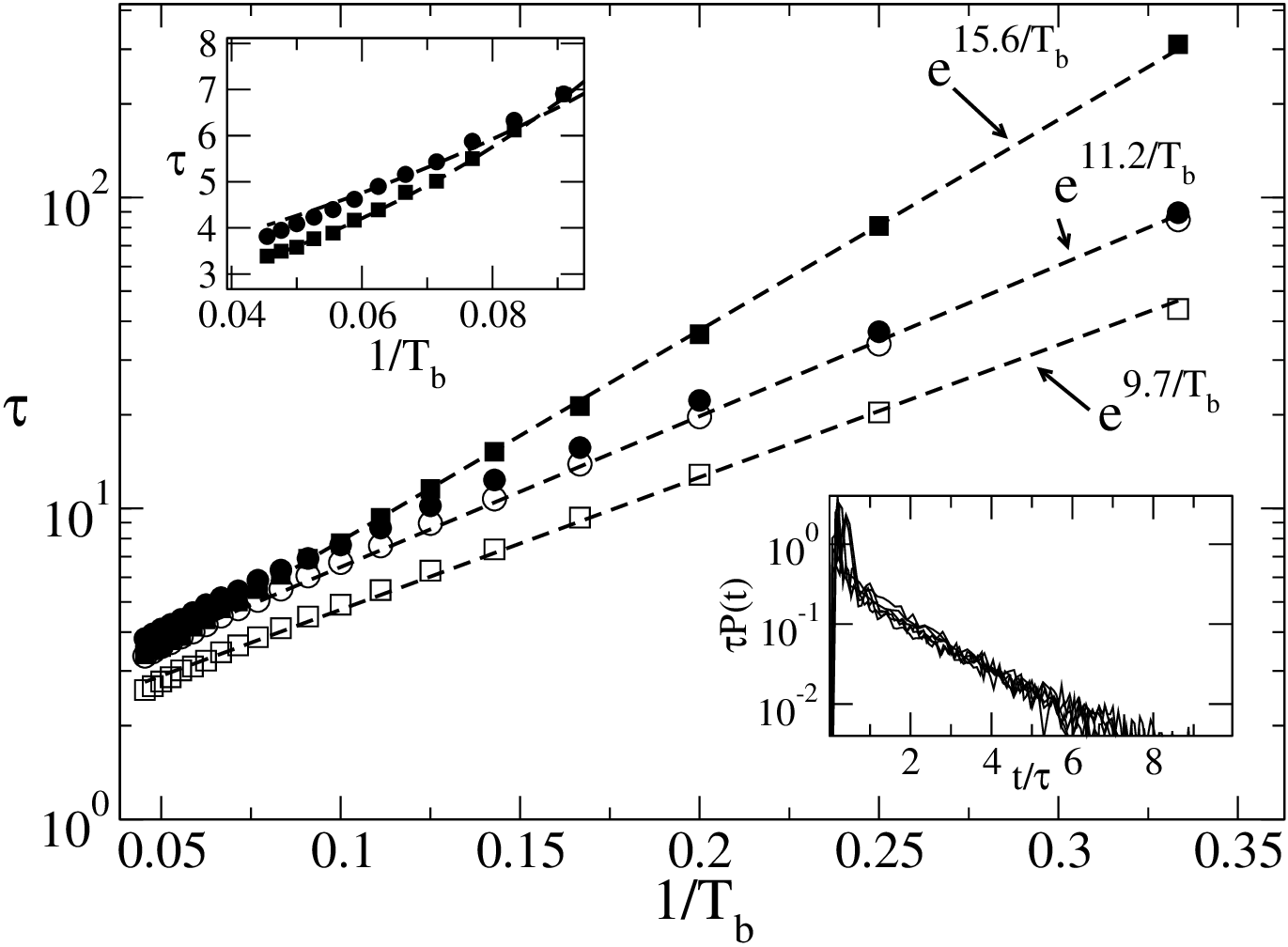}
\end{center}
\caption
{Arrhenius plot of mean escape times $\tau$. 
Open symbols refer to elastic case: the escape time 
is $\tau_1$ (circles) when a well
is singly occupied, $\tau_2$ (squares)
when a well is doubly occupied.
Full symbols, instead, correspond to the inelastic system ($r=0.9$).
Linear behaviour indicates the validity 
of Kramers theory with renormalized parameters and the slopes
agree with values obtained from Eq.~(\ref{newkramer}). 
Upper inset: enlargement of the crossover region. 
Lower inset: Data collapse of distribution of escape times for different
temperatures.
The arguments of the exponentials (dashed lines)
in the same figure have been obtained
by applying formula~(\ref{newkramer}). The overall
agreement between the above prediction 
and the values of $\tau_1$ and $\tau_2$ obtained by simulation
is rather good.} 
\label{fig_tempi}
\end{figure}

Two physical effects are present: the hard core repulsion and the 
inelasticity of collisions.
For the sake of clarity, let us begin the dicussion with the elastic 
system ($r=1$). In this case, we deal with an equilibrium system 
- the measured $T_2$ and $T_1$ coincide with the heat bath 
temperature $T_b$ \cite{distribuz} - and therefore,  
we expect that the most probable configuration minimizes the free 
energy. 
This configuration is constituted by a single particle in each well, 
and corresponds to the condition $\tau_1>\tau_2$. 
The results of simulations, shown in Fig.~\ref{fig_tempi}, 
verify this scenario. 
How do we quantify these escape times? 
As displayed in Fig.~\ref{fig_tempi}, $\tau_2$ and $\tau_1$
still follow the Arrhenius exponential behavior of Eq.~(\ref{kram}),
however, with a suitable paremeter renormalization:
\begin{equation} 
\label{newkramer}
\tau_k \approx \exp\big[\frac{W_k}{T_k}\big],
\end{equation}
where $k=(1,2)$ indicates single or double occupation, 
$W_1=\Delta U$ and $W_2=\Delta U - \delta U<\Delta U$.  
The correction $\delta U$ takes into account the effect of
the excluded volume repulsion: when two grains belong to the same well
their center of mass lies higher than if they were in separate
wells. This determines a reduction $\delta U = a(d/2)^2 + b/4(d/2)^4$ 
of the effective energy barrier and makes
$\tau_2<\tau_1$. This is a typical
correlation effect, because the repulsion renders less likely, with
respect to the non interacting case, the double occupancy of a well.
The smaller the ratio between the well width and the particle diameter,
the stronger the reduction of the escape time \cite{Tarazona}. 

\begin{figure}[htb]
\begin{center}
\includegraphics[clip=true,width=\columnwidth, keepaspectratio,angle=0]
{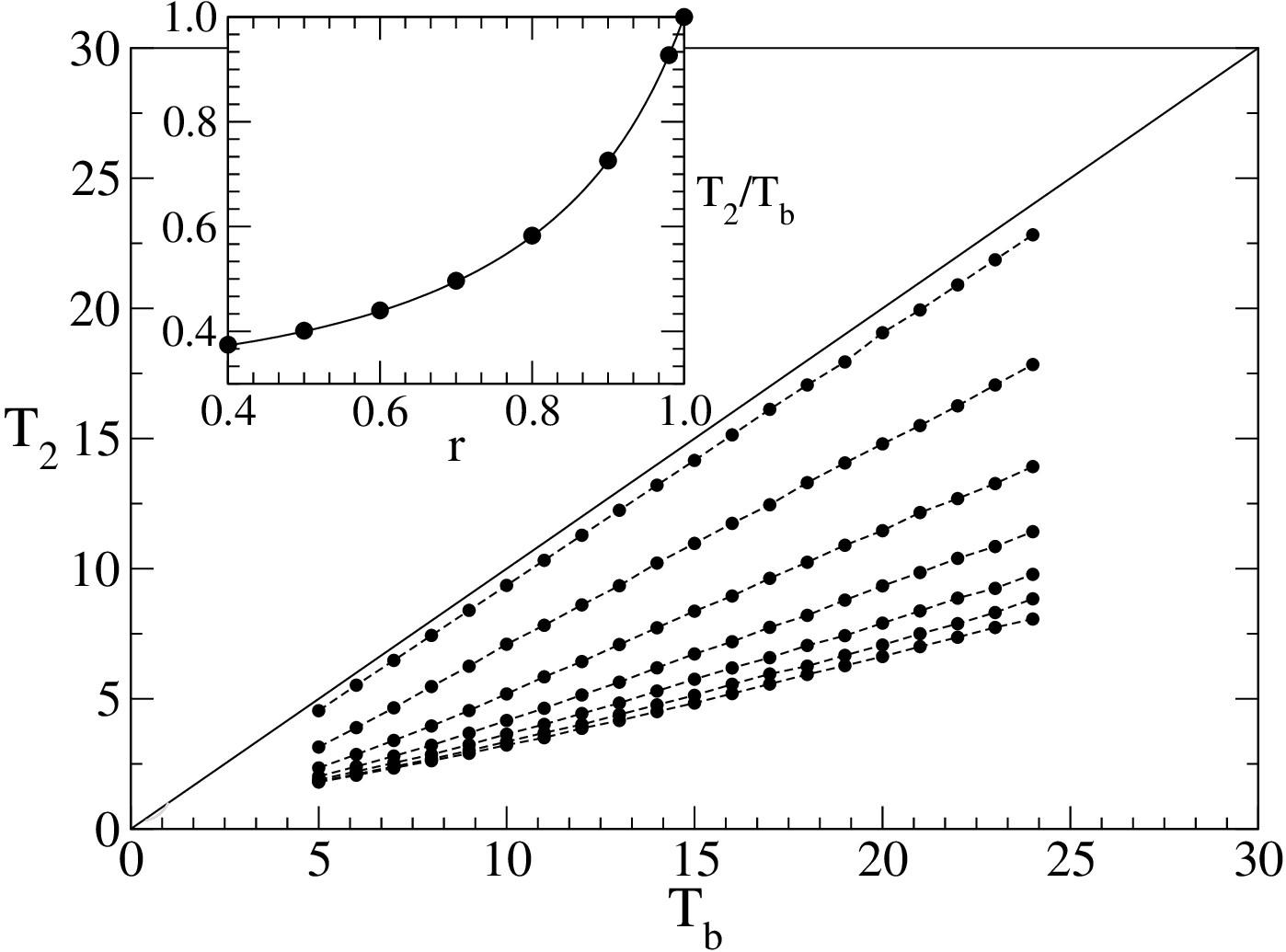}
\end{center}
\caption
{Plot of the temperature $T_2$ vs. $T_b$ for various choices of the
elasticity parameter $r$.
From top to bottom the curves indicate 
$r=0.98,0.9,0.8,0.7,0.6,0.5,0.4$. As $r$ grows toward the value $1$
the lines approach the diagonal (solid line) where $T_2=T_b$.
The inset shows the ratio $T_2/T_b$ obtained
by the slopes of $T_2$ vs. $T_b$ for the same values of $r$.
The continuous line indicates the theoretical estimate given by 
formula~(\ref{eq:umberto}).}
\label{fig_temperature}
\end{figure}

The above scenario changes in the presence of inelasticity ($r<1$)
because dissipation tends to promote the double occupation of a well. 
Thus, upon lowering the temperature $T_b$, we expect a crossover from 
the regime, where the double occupancy of a well 
is unfavoured (i.e. $\tau_2 < \tau_1$), to the regime, where 
particles spend most of the time together in the same well i.e. 
$\tau_2 >\tau_1$. 
In Fig.~\ref{fig_tempi}, for inelasticity $r=0.9$, this crossover is 
observed and occurs at $T_b \simeq 10.0$ (upper inset). 
This behavior is the analogue of that reported by several 
authors~\cite{Eggers,Brey}. 
The origin of the crossover lies on the fact that, 
in the inelastic system, temperatures $T_2$ and $T_1$ are no more equal 
to $T_b$ and furthermore $T_2 < T_1$. 
Thus, the mean lifetime of the clusterized and non clusterized regimes can 
be still described by expressions~(\ref{newkramer}), but now the temperature 
difference competes with the excluded volume correction, eventually 
leading to $\tau_2 > \tau_1$.
A simple argument can be used to estimate the shift of $T_2$ from $T_b$.
For moderate driving intensity, $T_1$ is nearly
equal to $T_b$, while $T_2$ is lower than $T_b$ by a factor which
depends on the inelasticity.  In Fig.~\ref{fig_temperature} we show
these temperatures as functions of $T_b$. It can be observed in Fig.~
\ref{fig_temperature} that $T_2$ varies linearly with $T_b$ and its slope is 
a decreasing function of the inelasticity $(1-r)$. 
A good estimate of temperature $T_2$ can be obtained by considering the two 
particles in a single
harmonic well $V(x)=\omega^2_{min} x^2/2$.  Balancing the power
dissipated by collisions $-(1-r^2)\nu\overline{(v_2-v_1)^2}/4$ and
viscous damping ($-2\gamma T_2$) with the power supplied by the
external source ($2\gamma T_b$) we arrive to the expression:
\begin{equation} \label{eq:umberto}
T_2=\frac{T_b}{1+\frac{\nu}{4\gamma}(1-r^2)} 
\end{equation}
where the collision frequency $\nu$ is estimated as
$\omega_{min}/\pi$ (the factor $2$ stems from the excluded volume
effect). We have assumed that the precollisional relative velocity
$\overline{(v_2-v_1)^2}\simeq (<v_1^2>+<v_2^2>)=2 T_2$.  
An improved value of $T_2$ results when
using the value of $\nu$ obtained from the simulation. The dependence of
the formula on the restitution coefficient is shown in the inset of 
Fig.~\ref{fig_temperature}. 

In the lower inset of Fig.~\ref{fig_tempi} 
we plot the probability distributions of
escape times $\tau_1$ and $\tau_2$ for several simulations. 
All the distributions are characterized by a peak at the
origin and an exponential tail. When rescaled to have the same
average, all the tails collapse to a single curve. Such exponential
tails are typical of the original Kramers
model for thermally activated barrier crossing. 

\begin{figure}[htb]
\begin{center}
\includegraphics[clip=true,width=\columnwidth, keepaspectratio,angle=0]
{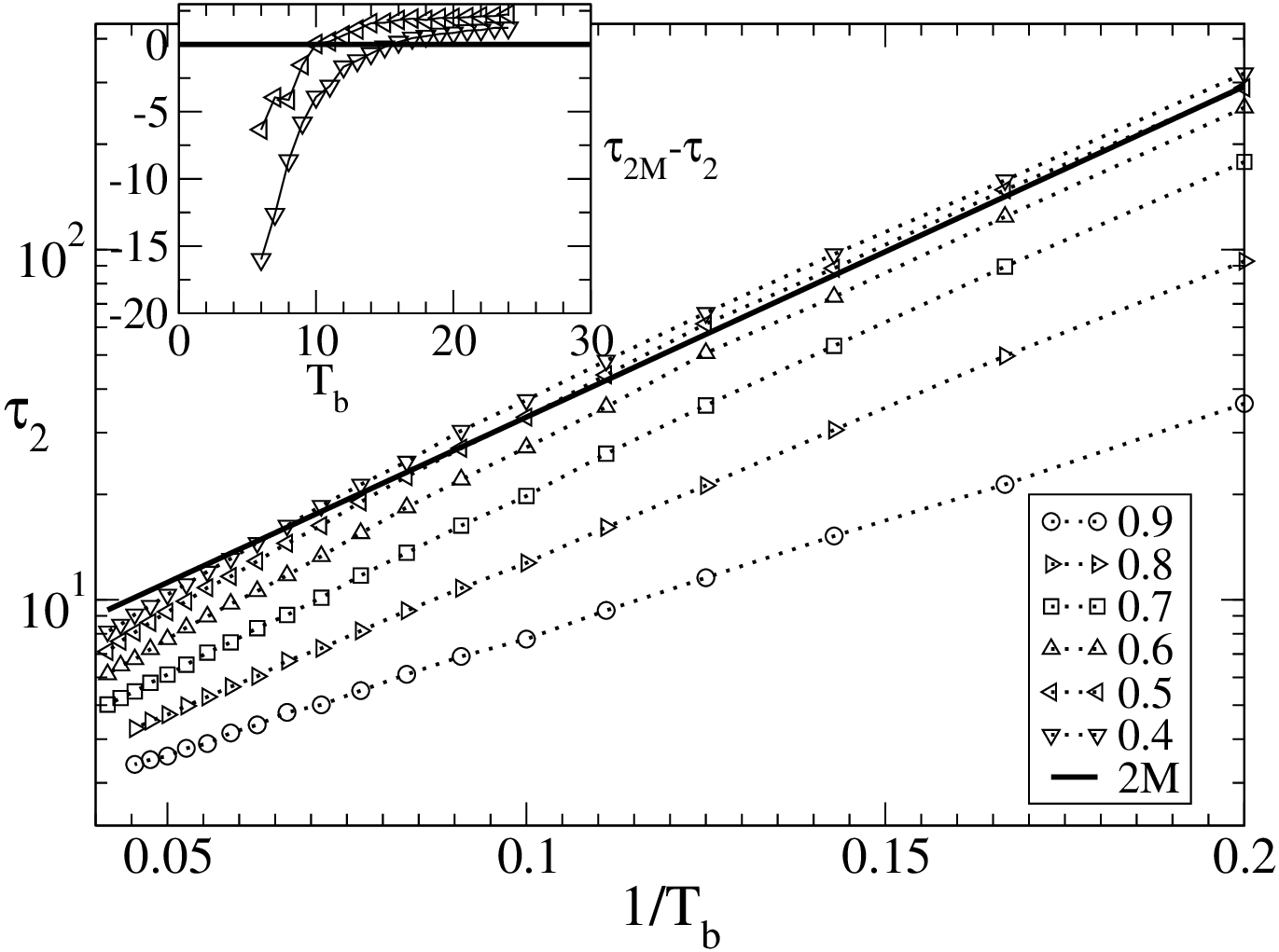}
\end{center}
\caption
{Arrhenius plot of the mean escape times $\tau_2$ versus $1/T_b$ 
for several values of $r$. Thick solid line indicates the escape time 
$\tau_m$ of single particle of mass $2M$ corresponding to simultaneous 
jumps. At about $T_b=10$ and $r=0.5$  we observe a crossover between 
$\tau_2$ and $\tau_m$ signalling the formation of a stable molecule.
The inset shows the  difference $\tau_m - \tau_2$ vs. $T_b$ to 
pinpoint the crossover where $\tau_m\simeq \tau_2$} 
\label{fig:collapse}
\end{figure}

A new non trivial phenomenon occurs
at low temperature and small restitution coefficient. 
The escapes from a doubly occupied well
become correlated, i.e. the transition takes place as a collective
motion of the two particles. In other words, when the inelasticity
is strong the relative motion of the two particles,
due to repeated collisions, becomes
frozen and they tend to form a ``molecule''. It is easy to show
(within an harmonic treatment of the potential) that the noise
acting on the center of mass coordinate corresponds to a reduced 
heat bath temperature $T_b/2$. We computed the exit time $\tau_m$
for the ``molecule'' and compared  it with the characteristic times
$\tau_2$ for several values of the inelasticity. As shown 
in Fig.~\ref{fig:collapse}, when the driving temperature decreases,
$\tau_2$ corresponding to a strongly inelastic system
($r=0.5$) exceeds $\tau_m$. This is the signature that the 
most probable evolution of a doubly
occupied state involves the simultaneous hopping 
of the two particles in the adjacent well, without breaking the
pair.
This freezing of the internal motion,
is related to the  problem 
of the dynamics of two randomly
accelerated particles on a line considerd in Refs.~\cite{Bray,Burkhardt}, 
or equivalently, to the problem of a single
particle moving on the half line ($x>0$)
in the presence of an inelastic (impenetrable)
wall. The authors predicted that at fixed driving intensity and for $r$
below a critical value, $r=r_c$, the
particle localises at the wall.
We argue that the localization mechanism is identical 
to that leading to the formation of the ``molecule'' in our model.

In summary,
whereas previous studies on compartimentalized systems were
based on coarse grained descriptions of granular gases,
our approach focuses on the statistical
description of particle motions by means of a characterization of their 
stochastic fluctuations.
The basic phenomenology illustrated
above, suggests
that after a suitable redefinition of parameters ($T_b$ replaced by
$T_2$ and $\Delta U$ by $W_2$) the present problem
can be mapped onto that of a single particle
hopping between two wells at different temperatures $T_1$
and $T_2$. We have shown how $T_2$ decreases with the inelasticity.
The local granular temperature is the relevant control parameter
in determining the 
direction of the energy flux (particles flow from hot places
to cold places). The activation rates, $\tau_2^{-1}$ and 
$\tau_1^{-1}$ display an Arrhenius dependence on the temperatures
$T_1$ and $T_2$, respectively. 
However, as $r$ decreases further,
a deeper scrutiny reveals a more complex scenario.
 The model displays a further ``transition'' at even low temperatures
because the two rods form a ``bound'' pair as a result of extremely frequent 
collisions. The escape from the wells can occur only ``in tandem''.

As a natural extension of the present model,
due to the continuous interest on transport properties on
rough surfaces, compartimentalized systems and ratchets,
one could examine the dynamics
of an ensemble of grains in a larger number of wells.
This would allow to determine explicitly the variation of
the granular temperature and of the transition rates 
with the occupation number of the wells. The latter are
key ingredients in hydrodynamic descriptions.

We are in debt to G.~Jug for his suggestions and remarks.
We acknowledge the support of the 
Cofin MIUR ``Fisica Statistica di Sistemi Classici e Quantistici''.


\begin{thebibliography}{99}

\bibitem{gases} Granular Gases, vol. {\bf 564} of Lectures Notes in 
Physics, T. P\"oschel
and S. Luding editors, Berlin Heidelberg, Springer-Verlag (2001). 

\bibitem{goldhirsch}
I. Goldhirsch and G. Zanetti, {\em Phys. Rev. Lett.} {\bf 70} 1619
(1993).

\bibitem{bal}
A. Baldassarri, U. Marini Bettolo Marconi and A.Puglisi,
{\em Phys. Rev.E}, {\bf 65}, 051301 (2002).

\bibitem{Ioepug} 
U.~Marini Bettolo Marconi and A.Puglisi
{\em Phys. Rev. E} {\bf 66}, 011301 (2002).

\bibitem{Nogauss} A.Kudrolli and J.~Henry {\em Phys. Rev. E} {\bf 62}, 
                  R1489 (2000).
\bibitem{jaeger} 
H.M. Jaeger, S.R. Nagel and R.P. Behringer, {\em
Rev. Mod. Phys.}  {\bf 68}, 1259 (1996) and references therein.

\bibitem{experiment}
H.J.~Schlichting and V. Nordmeier, {\em Math. Naturwiss. Unterr.}, {\bf 49}, 323 (1996) (in German).

\bibitem{Eggers}
J. Eggers, {\em Phys. Rev. Lett.} {\bf 83}, 5322 (1999).

\bibitem{Brey}
J. Javier Brey, F. Moreno, R. Garc\'{\i}a-Rojo, and
M. J. Ruiz-Montero, {\em Phys. Rev. E} {\bf 65}, 011305 (2002).

\bibitem{Lohse}
D. van der Meer, K. van der Weele, and D. Lohse, {\em
Phys. Rev. Lett.} {\bf 88}, 174302 (2002).

\bibitem{Kramers}
H. A. Kramers, {\em Physica (Utrecht)} {\bf 7}, 284 (1940).

\bibitem{Hangi} 
P. H{\"a}nngi, P. Talkner, M. Borkovec, Rev. Mod.Phys. {\bf 62}, 251
(1990).

\bibitem{tradition} 
The external drive is modeled via a stochastic force. See
A. Puglisi, V. Loreto, U. M. B. Marconi, A. Petri, and A. Vulpiani,
{\em Phys. Rev. Lett.} {\bf 81}, 3848 (1998); A. Puglisi, V. Loreto,
U. M. B. Marconi, and A. Vulpiani, {\em Phys. Rev. E} {\bf 59}, 5582
(1999).

\bibitem{distribuz}
We checked numerically that the velocity distributions
for the elastic system were gaussians of variance $T_b$.

\bibitem{Tarazona}
U. Marini Bettolo Marconi and P.Tarazona, J.Chem.Phys. 
{\bf 110}, 8032 (1999).

\bibitem{Bray}
S. J. Cornell, M. R. Swift, and A. J. Bray, {\em Phys. Rev. Lett.} 
{\bf 81}, 1142 (1998).

\bibitem{Burkhardt} T.W.~ Burkhardt, Phys. Rev. E {\bf 63}, 011111 (2001).

\end{thebibliography}
\end{document}